# MadQCI: a heterogeneous and scalable SDN QKD network deployed in production facilities.

V. Martin[1*], J.P. Brito[1], L. Ortiz[1], R.B. Mendez[1], J.S. Buruaga[1], R.J. Vicente[1], A. Sebastián-Lombraña[1], D. Rincon[2], F. Perez[2], C. Sanchez[2], M. Peev[3], H.H. Brunner[3], F. Fung[3], A. Poppe[4], F. Fröwis[4], A.J. Shields[5], R.I. Woodward[5], H. Griesser[6], S. Roehrich[7], F. De La Iglesia[8], C. Abellan[8], M. Hentschel[9], J.M. Rivas-Moscoso[10], A. Pastor[10], J. Folgueira[10] and D.R. Lopez[10]

[1]*Center for Computational Simulation, Universidad Politécnica de Madrid, Madrid, Spain*
[2]*IMDEA Software Institute, Madrid, Spain*
[3]*Munich Research Center, Huawei Technologies Duesseldorf GmbH, Munich, Germany*
[4]*Nutshell Quantum-Safe GmbH, Vienna, Austria*
[5]*Toshiba Europe Ltd., Cambridge, UK*
[6]*Adva Network Security GmbH. Berlin, Germany*
[7]*Rohde & Schwarz Cybersecurity GmbH, Germany*
[8]*Quside, Barcelona, Spain*
[9]*Austrian Institute of Technology, Vienna, Austria*
[10]*Telefónica gCTIO/I+D, Madrid, Spain*

## Abstract

Current quantum key distribution (QKD) networks focus almost exclusively on transporting secret keys with the highest possible rate. Consequently, they are built as mostly fixed, ad hoc, logically, and physically isolated infrastructures designed to avoid any penalty to the quantum channel. This architecture is neither scalable nor cost-effective and future, real-world deployments will differ considerably. The structure of the MadQCI QKD network presented here is based on disaggregated components and modern paradigms especially designed for flexibility, upgradability, and facilitating the integration of QKD in the security and telecommunications-networks ecosystem. These underlying ideas have been tested by deploying many QKD systems from several manufacturers in a real-world, multi-tenant telecommunications network, installed in production facilities and sharing the infrastructure with commercial traffic. Different technologies have been used in different links to address the variety of situations and needs that arise in real networks, exploring a wide range of possibilities. Finally, a set of realistic use cases have been implemented to demonstrate the validity and performance of the network. The testing took place during a period close to three years, where most of the nodes were continuously active.

## Introduction and Overview

QKD technology, the ability to grow a secret key between two partners to a practically unlimited size and with bounded information leakage, has been steadily advancing since its first implementation in 1989 [1]. What was then a few tens of centimeters is now about a thousand kilometers [2], [3] in fiber links and ground-to-satellite connections [4]. However, QKD is a demanding and still evolving technology that deals with signals at the lowest possible intensity and this imposes hard physical limits in terms of maximum absorption or tolerated noise. In the absence of quantum repeaters [5, 6], point-to-point QKD links have an ultimately limited reach.

---

* Vicente.martin@upm.es

To overcome the QKD limits and serve as many users as possible in practical applications, QKD networks have been built over the last two decades [7] and implemented all over the world to demonstrate different objectives [6] [8] [9] [10] [11]. These deployments are QKD-centric, meaning that the architecture is designed to maximize key rate while avoiding problems with the quantum channel. Additional ad hoc infrastructures, separated from the normal telecommunications network and following their own operational rules, are specifically built for this purpose. These networks, although very important and showing great advances, are also very costly and not compatible with the software and hardware architecture that supports typical telecommunications networks. None of these QKD networks are offering services to users in the same way as a standard telecommunications network does, heavily penalizing their commercialization. A tighter integration in the day-to-day telecommunications and security ecosystems, allowing for infrastructure reuse and the provision of services in a cost-effective way, is needed to grow QKD to a mainstream technology that will benefit our society.

However, the path to a QKD network that shares and integrates well with classical infrastructure, including management and operational procedures, is yet to be found. Modern and flexible networking paradigms were tested in the field [12] but more extensive research on the different technologies and their interaction is still needed.

In this paper we present a highly heterogeneous quantum network fully integrated within a commercial optical telecommunication one. It has been deployed in production networks running commercial services. It is also based on software defined networking (SDN) [13] since this paradigm has been demonstrated to be flexible enough to support QKD devices within a classical network architecture [12]. The quantum part of the current version of the Madrid network, which we call MadQCI (Madrid Quantum Communications Infrastructure), is composed of 28 QKD modules (emitters and receivers) and a QRNG service. The devices are provided by five different manufacturers and installed in 9 production sites, separated between 1.9 and 33.1 km. Part of the network is switched, the pairing of devices is flexible and on demand. Whenever the maximum tolerated losses allow it, direct quantum links can be dynamically established between a set of the nodes while bypassing some of them. To demonstrate resilience, several links are served by multiple quantum channels using devices with different technologies and provenance. This provides an additional degree of redundancy and security also by reducing the dependence on a single manufacturer. The dynamical capacity introduced by the switching reduces the number of trusted nodes and increases the total number of possible direct quantum links to 45. Moreover, the QKD devices are in different private networks, one per manufacturer, such that they are logically disconnected to further increase resiliency and security. The optical links between nodes are just a pair of strands of optical fiber, which carry quantum and classical communications as well as the service signals. In some links, several quantum channels -and the corresponding service ones- from different QKD devices using different technologies and in dissimilar configurations shared the same fiber. To allow classical and quantum signals over the same physical infrastructure and not to risk breaking the very strict service level agreements of the classical communications providers, a range of solutions has been deployed in the different links.

The whole network is managed and operated using the SDN paradigm, applying standards developed in the European Telecommunications Standardization Institute (ETSI) [14, 15, 16] for QKD. The efforts described in this paper have also helped in refining these standards, since this is the first time that they are used in such a complex network, with a physical infrastructure spanning two different network providers: Telefónica, the largest operator in Spain, and RedIMadrid, the network provider for the research and educational community in the Madrid region. Border nodes, served by two different quantum links, connect both networks and allowed the creation of secure links from any to any node, even when these belong to different domains. Finally, a QRNG service was also integrated as a source of entropy to be used in a set of applications.

MadQCI is simultaneously operating many use cases related to different sectors. All associated classical communications needed to keep the network and use cases running use standard equipment without any special adaptation and share the same infrastructure. This implies quantum/classical coexistence at the link level and, in most cases, sharing the same physical media. The only modification is in the encryptors used to cipher the communications. These are also commercial devices, but the firmware has been adapted to refresh the keys from the quantum network using ETSI standards. Encryption can be done at OSI network levels 1, 2, and 3 depending on the specific requirements of the use cases in terms of latencies or interfaces, so that the network services are truly transparent to the applications. This modular architecture facilitates a better integration in the security ecosystem and joint use of QKD and conventional, computational-complexity based, cryptography in a step towards crypto agility, preempting the transition of current networks towards quantum-safe ones.

The MadQCI network has been continuously operating in different stages over the last years. Some segments were running without interruption - except for maintenance, new software installation, etc. – close to three years. Most of the devices were operating over the last year and two links were in production during the last three months.

To the best of our knowledge, this makes MadQCI the largest and longest running QKD network in Europe. The demonstrated architecture was developed as a blue-print for future, forward looking deployments. This includes complex scenarios for exploring and demonstrating the maturity level of the technology and tackle ambitious projects such as EuroQCI, the ten years program to build a pan-European Quantum Communications Infrastructure.

The paper first describes the logical architecture, physical devices, and optical layout of the network, delving in its integration capabilities and other significant aspects such as dynamical switching, and finally, for the sake of completeness, describes a sample of the applications that were tested, highlighting some of the specific metrics, before the concluding remarks.

## Architecture

Many QKD networks have already been built [9], but they mostly share the same characteristic: they concentrate, almost exclusively, on maximizing the key throughput. To achieve this, their architecture has been tuned to minimize any disturbance in the quantum channel. Thus, the use of dark fiber for the quantum channel has been prevalent. In fact, most of these networks can be seen as an ad-hoc, separate network, built solely for quantum purposes that use any classical network available for the associated classical communications. This approach requires to build a specific infrastructure just for QKD. While this might be adequate for early adopters or research motivated but temporally limited testbeds, it presents several challenges for its wide-spread usage. Not reusing or not sharing existing infrastructure is very costly and demands a large investment up-front. It is not only about optical fiber, but also about additional management costs and suboptimal use of the network, dealing with proprietary interfaces, specialized maintenance, and, in general, a lack of flexibility and interoperability. Such designs inhibit a scale-up of the network and adding systems in a multivendor infrastructure.

To avoid these problems, the network presented here was built following a completely different approach. Its architecture follows the SDN paradigm [13], designed to increase the flexibility and shorten the times for deployment and maintenance. Standards and well-known tools in the telecommunications industry were extensively used to facilitate integration and adoption.

The fundamental concept of SDN is the separation between the control and data planes. In an SDN environment, the data plane, considered as the set of data and functionalities provided by the network to ensure traffic from source to destination, is bounded to dedicated elements (*forwarding functions* or *devices*). In an SDN based QKD network, these functionalities include the key transport capability. The control and management tasks are mediated by the SDN controller that offers a programming interface to the control of network behavior. This includes the reaction of the SDN based QKD network to failures or malfunctions, e.g., detected security breaches. Moreover, the mechanisms to export the capabilities offered by the forwarding functions to the control plane are standardized. This results in a very flexible and powerful infrastructure that can incorporate new devices and technologies, facilitating interoperability and a much quicker deployment of networks and services, compared to previous paradigms. At the same time, the network management is also simplified since the whole network can be viewed and managed through the controller. This is what makes the SDN paradigm so popular among telecommunication companies.

From a QKD point of view, the SDN controller can obtain information on the devices installed in the network and their characteristics. QKD systems are treated as network devices that export their capabilities to the network. Note here that we are referring to the functionality, not to security-related issues, like the secret key itself that remains unknown to the controller. Depending on how much functionality the QKD module exports to the network, the integration can be as simple as commands to start, stop, and resynchronize the QKD module or as sophisticated as, e.g., to manage a single sender/receiver as an endpoint of many receivers/emitters in a one-to-many/many-to-one/many-to-many configuration. Whereas the controller does not access secure data, it knows the key requirements of different applications and can set the routes to forward keys to fulfil given service level agreements. This includes pre-emptive key storage, optical-power management in the fibers, and resource-aware optical-route planning to optimize the network for optimal performance. Dynamic optical-route planning with switches to create different sender/receiver pairs or wavelength selection is also possible as well as setting quality-of-service (QoS) parameters for specific users. Other, more sophisticated, control and management tasks, such as doing network slicing, multitenancy or creating a border node between two networks belonging to different operators can also be performed. The functionality of creating a border node is crucial to substantially grow a network in a cost-efficient way, something that we have demonstrated in MadQCI.

The MadQCI design is shown in Figure 1. The basic node scheme follows the approach of a software defined QKD node (SDQKDN) that was used in a previous trial [12] and in contributions to SDN-QKD standards [16].

The network nodes can have several QKD modules installed. They interact through interfaces with a set of disaggregated software components implementing clearly defined functions [17]. This allows for the design of interfaces in a vendor-independent manner for structured communications. The interfaces were implemented with well-known tools and adhering to standards in telecommunications, which also helps to create confidence in the technology. The approach is scalable and flexible: it can be extended to many nodes, be used to increase the capacity of each node, support a variety of QKD technologies and even to support new services beyond QKD. The components within a node are:

- Local Key Management System (LKMS): it collects the keys from the QKD modules and serves the applications; indexes and stores the generated keys, manages their lifecycle, and keeps track of the key-generation peers; provides information on key availability to the SDN controller through the SDN agent and the keys to be forwarded when needed. Below we use the general term Key Management System (KMS) to address the functionalities of the set of all LKMSs.
- Forwarding Module: it is in charge of the key transport between nodes using the shared keys created by the QKD module pairs. In contrast to typical implementations, this functionality is here separated from the LKMS, since key routing is not a part of KMS duties as defined in, e.g., the NIST SP 800 document series. This facilitates the integration into the standard security ecosystem.
- SDN Agent: SDN controller counterpart in each node that connects the controller with all the components within the node. Note that security-sensitive information is not available to the control mechanism.
- QKD Module: the quantum sender/receiver itself, which continuously generates the keys. In general, there are three channels associated to it: the quantum channel, a service channel, needed to stabilize the quantum channel (possibly integrated with the former), and the classical key-distillation channel.
- Application: any entity, inside the SDQKDN security perimeter, requesting QKD keys from the LKMS. The applications might be external, e.g., an end-user application, a hardware security module (HSM), a virtual network function, or internal with respect to the key distribution functionality, e.g., authentication, virtual link management, or key transport. The applications use the application interface implemented in the LKMS to obtain the key material.

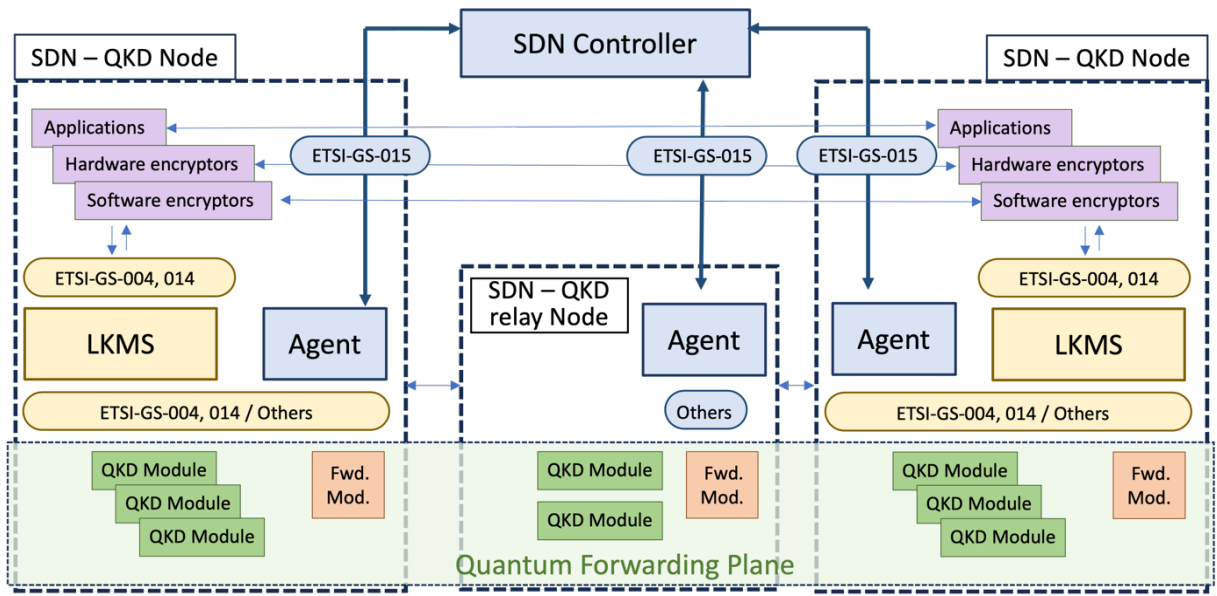

*Figure 1: SDN-QKD architecture. The different components (see text for details) in a node are shown within the dashed line boxes. The rounded boxes with numbers indicate the corresponding ETSI ISG QKD group specification used as an interface between the components above and below or in the connector line. We depict three nodes, each node connects with other ones in the network using a set of quantum and classical channels for quantum signal transmission, the service channels to keep the quantum links working, the key distillation and user data channels. Other classical channels can be established between the encryptors or software applications running in the node. Not all nodes serve user applications since they can be used as trusted relays with the only purpose of extending the QKD reach. For this reason, relay nodes have no KMS functionality. Key routing, which cannot be considered a KMS functionality, is managed by the Forwarding Module. The middle node is an example of a relay node. We encapsulate all the functionality needed to create the secure key transport capability that characterizes a QKD network in the lowest layer, the quantum forwarding plane, analogous to the data forwarding plane in standard SDN networks. It includes the QKD modules together with a forwarding module that implements a secure key forwarding mechanisms.*

This set of components is sufficient to implement all the functionality required by a QKD network. They are also flexible enough to cope with new applications and substantial enough to represent a possible target for standardization. Following the SDN paradigm, the node communicates with the (logically) centralized SDN controller that implements all necessary logic and interfaces to control the network. The controller creates the logical and physical connections necessary for sharing a key end to end. It also provides the interface to the network management system, allowing advanced functionalities, e.g., setting QoS parameters for different users, the orchestration of several networks, the creation of large multi-domain, multi-tenant networks. A sought-after effect of this disaggregated approach, open-standard interfaces, and communications is to allow for vendor independence and to reuse as much of the existing communications technology as possible. This is done again with the objective to create confidence and allow an as easy as possible integration of QKD technology in the communications and security ecosystem.

The nodes of MadQCI were deployed in the network as shown in Fig. 2. This network was not created ad hoc but uses a pre-existing production network that provides services to commercial customers. It is important to remark that all the installed QKD systems were located in production facilities under typical, carrier-grade, working conditions. No system was operated in a lab environment except the experimental link that was connected remotely during a limited time. In total, 28 QKD modules (counting emitters and receivers, 26 on-site and 2 remote) using different QKD technologies and protocols were installed (see Table 1) in 9 nodes of the Telefónica and RedIMadrid production networks. Both networks were connected through special border nodes. The length of the links ranged between 1.9 and 33 km (optical losses between 2.0 and 14.3 dB in the C-band), covering the Madrid metropolitan and suburban area. Except for one link (link 3 in Fig. 2) that uses two pairs, all nodes are connected through a single pair of optical fibers that carry all quantum and classical signals. No ad hoc fiber was deployed specifically for the quantum channel. Coexistence of the quantum and classical channels was a must, as well as compatibility with standard optical transport (OTN) equipment and cryptographic appliances. For maximum flexibility and transparency to the application layer, encryption can be done at OSI levels 1, 2, and 3. Level-3 encryption (IPsec) was done via software implementation of AES, as well as one-time pad encryption. Off-the-shelf OTN and network encryptors from ADVA and Rohde & Schwarz were used. The R&S firmware of the level-2 encryptor was adapted to extract QKD keys with the ETSI GS QKD 004 [14] standard and use them to cypher communication using AES at rates up to 40 Gbps. Encryption at level-1 was done using the ADVA encryptors, also modified to accept the key using the GS QKD 014 [15] standard.

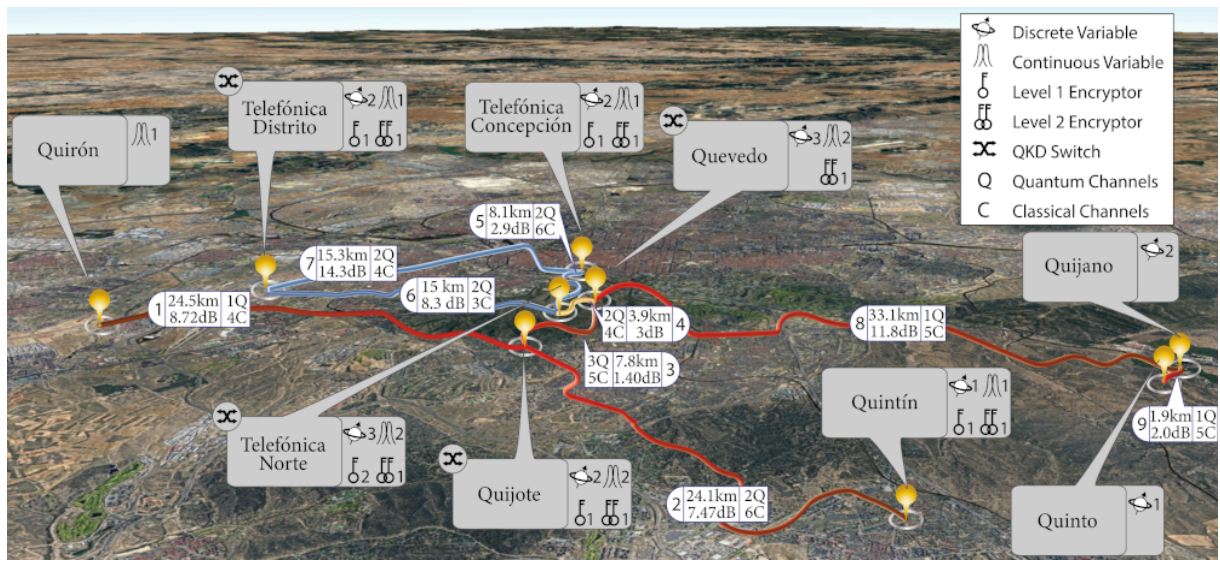

*Figure 2: Physical layout of Madrid Quantum Communications Infrastructure. It is composed of 9 sites in two production networks with 26 QKD modules on-site and 2 experimental ones installed remotely (not depicted). White labels refer to information about the links. Grey labels refer to nodes. The links in red belong to RedIMadrid, the network provider for all research and educational institutions in Madrid. The ones in blue belong to Telefónica. A border link (yellow) connects the two networks. All nodes are in production facilities, sharing the location with classical equipment and optical fibers carrying classical communications. Each link is marked with a label (number in the rounded part of a white label) for further reference together with the optical losses, physical distance and number of quantum and classical channels on the same fiber. Each node is tagged with the number and type of QKD modules installed, the OSI level of the classical hardware security module (Level 3 and OTP was always available through software) and whether there is an optical switch for the quantum channel installed. The switches allow the creation of direct quantum channels among non-contiguous nodes, raising the total number of possible connections to 45, up from the 9 possible ones in the physical topology depicted above (see Table 1 for additional details).*

An important aspect of the network is that the optical connection infrastructure is not static. Several all-optical switches, managed through the SDN controller, were installed. Specifically, the quantum channels were not static and could be established with different endpoints. Again, we did this using standard telecommunications technology, both at the hardware and software level (OADM modules built from standard, readily available components, and Transport API) to demonstrate compatibility. In this way we had many more direct connections (i.e., with an uninterrupted quantum channel) than those strictly linking one node to its nearest neighbor. A total of 45 compatible direct connections were possible with optical losses low enough to create QKD keys, which is substantially more than if it was a traditional, fixed QKD network physically laid out as in Fig. 2. The SDN controller, together with the LKMS and key forwarding module, can distribute end-to-end keys between any two nodes in the network, no matter the vendors or combination of these in the connecting path. The controller can also regulate the key provisioning, thus supporting QoS constraints and making the network more resilient to connectivity failures.

## QKD Systems

To show that many QKD devices, not just in quantity but also in type, can interoperate in a network during considerable periods of time is a key requirement for operators before deploying the technology in the real world. This heterogeneity was specifically sought after when designing the network. In this section we describe the deployed QKD technology.

The 10 Huawei CV modules were continuously available, except for servicing, to run the use cases close to three years. The 8 ID Quantique DV modules were running most of the time, in periods of months, during the last two years, while 4 Toshiba DV modules were installed and running continuously during the last year and other 4 additional ones during the last three months, when the network was running in its full configuration. The two, experimental, QKD modules from AIT were connected, albeit remotely, to the network during shorter time periods to demonstrate how to adapt new devices easily. However, no use cases were run using them. The distance between emitter and receiver in the remote link was 4.3 km. To further show the flexibility of the approach, a QRNG provided by QuSide was integrated in the network. The service could be used to produce on-demand, high-quality random numbers from an independent vendor for, e.g., key-generation purposes. A detailed list of the modules and their main characteristics is collected in Table 1.

*Table 1: QKD systems installed in each link and their main characteristics: Type of technology (Continuous/Discrete Variables), optical band for the quantum channel, average secret key rate throughput (and QBER for DV systems) in that particular link and number of quantum and classical channels sharing the link. See Fig. 2 or 3-a) for the link number and additional information. Note that data here is collected per link (white boxes in Fig. 2). The number of quantum and classical channels are quoted only for the physical links directly connecting the nodes, since the combinations of links might carry different number of quantum and classical in each of the links. Note also the different combinations arising from the ability to use different wavelengths and different propagation directions of the quantum channel (e.g., in Link 3). 6 back-to-back links, which are irrelevant from a network point of view, are not listed.*

| Link number (or combination of links) | Manufacturer | Type | Optical band | Average skr throughput (kbps) | QBER (%) | #quantum channels | #classical channels |
|---|---|---|---|---|---|---|---|
| 1 | HWDU | CV | C-37 | 3.3 | - | 1 | 4 |
| 2 | HWDU | CV | C-37 | 5.5 | - | 2 | 6 |
|   | HWDU | CV | C-38 | 6.2 | - |   |   |
|   | ID Quantique | DV | C-34 | 1.9 | 2.4 |   |   |
| 3 | ID Quantique | DV | O | 2.04 | 2.2 | 3 | 5 |
| 3 (Quijote->Quevedo) | HWDU | CV | C-37 | 4. | - | 3 | 5 |
|   | HWDU | CV | C-38 | 2.25 | - |   |   |
| 3 (Quevedo->Quijote) | HWDU | CV | C-37 | 11 | - | 3 | 5 |
|   | HWDU | CV | C-38 | 7.8 | - |   |   |
| 4 | ID Quantique | DV | O | 1.4 | 4.2 | 2 | 4 |
| 4 (Quevedo->Norte) | HWDU | CV | C-37 | 8.7 | - | 2 | 4 |
|   | HWDU | CV | C-38 | 12 | - |   |   |
| 4 (Norte->Quevedo) | HWDU | CV | C-37 | 9 |   | 2 | 4 |
|   | HWDU | CV | C-38 | 6.2 |   |   |   |
| 5 | HWDU | CV | C-34 | 0.09 | - | 2 | 6 |
|   | Toshiba | DV | O | 1039.9 | 3.4 |   |   |
| 6 | HWDU | CV | C-37 | 8.1 | - | 2 | 3 |
|   | HWDU | CV | C-38 | 8.4 | - |   |   |
|   | ID Quantique | DV | C-32 | 1.5 | 3.3 |   |   |
| 7 | HWDU | CV | C-37 | 7.4 | - | 2 | 3 |
|   | Toshiba | DV | O | 242.3 | 4.0 |   |   |
| 8 | Toshiba | DV | O | 37.2 | 3 | 1 | 5 |
| 9 | Toshiba | DV | O | 2857.1 | 2.5 | 1 | 5 |
| 6+7 | HWDU | CV | C-37 | 0.11 | - |   |   |
| 4 (Quevedo->Norte) +7 | HWDU | CV | C-37 | 2.4 |   |   |   |
| 3 (Quijote->Quevedo) + 4 (Quevedo->Norte) | HWDU | CV | C-37 | 7.1 | - |   |   |
|   | HWDU | CV | C-38 | 5.2 | - |   |   |
| 2 + 3 (Quijote->Quevedo) + 4 (Quevedo->Norte) | HWDU | CV | C-37 | 0.07 | - |   |   |
|   | HWDU | CV | C-38 | 0.01 | - |   |   |
| 4 (Norte->Quevedo) + 6 | HWDU | CV | C-37 | 4.3 | - |   |   |
|   | HWDU | CV | C-38 | 4.3 | - |   |   |
| 2 + 3 (Quevedo->Quijote) + 4 (Norte->Quevedo) | HWDU | CV | C-37 | 1.8 | - |   |   |
|   | HWDU | CV | C-38 | 1.7 | - |   |   |
| 3 (Quijote->Quevedo) + 4 (Quevedo->Norte) + 6 | HWDU | CV | C-37 | 0.04 | - |   |   |
|   | HWDU | CV | C-38 | 0.04 | - |   |   |
| 3 (Quevedo->Quijote) + 4 (Norte->Quevedo) | HWDU | CV | C-37 | 6.0 | - |   |   |
|   | HWDU | CV | C-38 | 6.6 | - |   |   |
| 1 + 3 (Quevedo->Quijote) | HWDU | CV | C-37 | 0.07 | - |   |   |
| remote | AIT | CV | C | 14 |   | 1 | 0 |
| remote | QuSide | QRNG | N/A | N/A QRNG speed: 4 Gbps. |   | N/A | N/A |

## AIT CV-QKD modules

The QKD modules developed by AIT provide a fully integrated CV system housed in two 19" inserts (Alice & Bob). It uses a QPSK constellation with 100 MBaud symbol rate for the quantum states and a polarization multiplexed pilot tone as phase reference. All digital driving signals are generated on an FPGA platform in the transmitter device, ready for receiving random numbers from a physical QRNG device. Auxiliary signals for packet triggering and clock synchronization are wavelength-multiplexed directly on the quantum channel, enabling a true single-fiber operation. In the receiver a true local oscillator is employed for heterodyne coherent detection, together with a 90° optical hybrid and two balanced receivers each for the quantum signals and the pilot. Automatic feedback loops for polarization control and laser frequency stabilization are implemented to ensure a stable long-term operation. Separate computers on either side perform the system hardware control, digital signal pre-processing and real-time post-processing. Pre-processing consists of down-sampling, frequency offset correction, phase correction, and parameter estimation. Parameter estimation is required for several algorithms in the post-processing stack, including calibration measurements of the thermal system noise and optical shot noise, an estimation of the SNR of the quantum states as well as their excess noise. The post-processing pipeline is instantiated on either side and communicates on an authenticated classical channel. The first step is an optional post-selection algorithm, which may improve the overall performance by trading the raw key rate for a better SNR. Next is information reconciliation by means of LDPC error correction followed by a confirmation of its success. Finally, all information leaked to a potential eavesdropper during physical key exchange and classical post-processing is rendered useless by performing privacy amplification. Here, an upper bound of the leaked information is calculated following the theoretical security proof. The final key is then reduced by this amount and all classically disclosed information (e.g., during information reconciliation) in a hashing algorithm to form the secure key. In the demonstration within MadQCI, whose purpose was to show the capability to easily integrate new, even experimental devices, in a running QKD network, those keys were forwarded to a KMS remotely over a raw TCP connection.

### Huawei Technologies Duesseldorf GmbH (HWDU), CV-QKD prototypes

As a partner in CiViQ and in the OpenQKD open calls, HWDU supplied 10 CV-QKD modules (5 senders and 5 receivers) to MadQCI. These modules are flexible, any sender could interoperate with any receiver, whereby the quantum channel between them could be optically switched, and moreover, as the lasers in both the senders and the receivers are tunable, wavelength switching in a broad range was feasible. Considering different possible paths between the 7 module locations, 36 different QKD links were supported, far more than the 5 links that would be available in a static configuration.

Each pair of the low-noise and low-complexity modules is generating key by means of a Gaussian-like modulation of coherent states and heterodyne coherent detection. It is to be noted that we are approximating true Gaussian modulation, an approach that is known to yield insignificant difference in key rate as a function of attenuation, compared to the theoretical case of analytic modulation [18]. The symbol rate is 12.5 MBd. Both, sender and receiver modules, feature phase and polarization diversity. Further, in-band synchronization is supported, whereby only one dense wave-length division multiplexing (DWDM) channel in the C-band in one direction is needed for the QKD operation. Additionally, a bidirectional, standard internet protocol (IP)-based post-processing link, that can operate on any existing network infrastructure, is required. The sender modules can transmit any chosen value between 0.0004 and 40 photons per symbol on average in the quantum band. Optical samples are generated, detected, and transformed from and to symbols utilizing appropriate DSP algorithms, running predominantly on an FPGA based SoC. Post-processing of symbols follows the traditional steps for CV-QKD, whereby error correction runs with a single fixed-rate code, which supports a signal-to-noise ratio (SNR) down to -19.5 dB. This is a receiver sensitivity of approximately -105dBm with 2.5 dB of receiver loss and heterodyne detection. Excess noise powers smaller than 50 dB below the shot noise can be detected. With trusted detector noise and an inherent system noise as low as 0.15 mSNU receiver side, a single sender-receiver pair supports up to 23 dB of channel loss.

### ID Quantique's Cerberis 3

In the framework of the OpenQKD project, ID Quantique supplied in total 16 pairs of QKD modules to different testbeds across Europe supporting different use-cases during time spans ranging from few months to more than one year. MadQCI received four fixed pairs connecting each transmitter to its corresponding receiver. Two links have the quantum channel at the O-band (1310 nm) and other two in the C-band, at ITU channels 32 and 34. The 1310 nm systems were delivered with additional built-in spectral filters to allow multiplexing of the quantum channel with classical channels in the C band. The systems are Cerberis 3 products, which were the predecessor of the current product Cerberis XG.

The implemented QKD protocol of the Cerberis 3 is the Coherent One Way (COW) with time-bin encoding [19]. COW does not require phase randomisation between consecutive pulses and uses a simpler decoy mechanism. This reduces the complexity of the optics for preparation and detection. The MadQCI devices implemented the last version of the SW, which include the countermeasure to the recently discovered attacks on the COW protocol. This countermeasure has little impact on the performance of the system. However, the typical real-world conditions, including temperature changes depending on time and local environments, caused larger than expected fluctuations in its performance. The new generation of QKD products is designed to deal with non-optimized cooling situations. The production environment of MadQCI enabled the optimization of stable operation.

The Cerberis 3 product includes a complete software suite to build up stand-alone trusted node QKD networks composed of ID Quantique devices. The network control and management are then run on an external server. However, to demonstrate and test the MadQCI interoperable setup, the Cerberis 3 were integrated as simple point-to-point links, and keys were requested via ETSI GS QKD 014 by the local KMS module and managed by the global, SDN-aware KMS, enabling network-wide end-to-end secure key transport. This demonstrates the flexibility which operators have already today to build up complex QKD networks.

### Toshiba QKD Devices

The 4 QKD systems (8 modules in total) from Toshiba implement the T12 protocol [20] which is an optimized version of BB84 with weak coherent pulses and decoy states. Phase-encoded quantum states at 1 GHz clock rate are generated by a gain-switched laser followed by an asymmetric Mach-Zehnder interferometer in the QKD transmitter [21]. In the receiver, self-differencing avalanche photodiodes (APDs) are used to measure the received states. The system fits into a standard 19-inch data centre rack and occupies 3 rack-units (3U) per node.

The Toshiba QKD devices in MadQCI all used a quantum channel wavelength of 1310 nm, optimized for supporting co-propagation of quantum and classical signals by maximizing the spectral separation between multiplexed quantum and classical channels. In addition, the systems include high-extinction narrow-band spectral filtering and time-domain gating of the APDs to further isolate the quantum channel from co-propagating classical / Raman-scattered light with maximal signal-to-noise ratio. This enables, for example, co-propagation of quantum light with over 60 x 100 Gbps DWDM channels in the C-band over 50 km, while still maintaining >100 kbps QKD secure key generation rates.

The QKD systems also include automatic self-optimization routines to dynamically adjust various optical parameters (e.g., polarization, phase, timing delays etc.) to maintain long-term maximal performance on each communication link [21]. Add/drop multiplexing hardware is also included in the QKD unit to multiplex the quantum channel, QKD service channels and any auxiliary data channels onto the communication link. Finally, QKD-generated keys are exported to the MadQCI network KMS at each node using the standardized ETSI GS QKD 014 interface.

### Quside QRNG

As an additional service in the network to be used in the MadQCI use cases, a Quantum Random Number Generator was made available. The device consists of a Quantum Entropy Source (QES) based on the proprietary phase diffusion technology [22] together with the firmware required to calibrate, control, monitor and provision of the entropy generated in the QES to an operating system. It is able to produce very high quality random bits at a speed up to 4 gbps. The QES and control electronics are integrated into a commercial PCIe card that was installed in a regular server located in the Telefónica network. Subsequently the QKD systems were used to transport the key from the QRNG through the network using one-time-pad encryption for the use cases where it was needed. The PCIe card is attached to a virtual machine where the corresponding drivers and libraries are installed and a server leveraging the libraries was deployed to provide entropy on demand through a simple REST interface. The key relay service present in the network requested entropy from this server for encryption key generation.

### Optical Transport and Encryptors

As mentioned above, encryptors modified to fetch keys from the local key management systems for the encryption were used. For the R&S SITLine ETH layer 2 Ethernet encryptors it was important that they could continue to operate in their view of the classical (possibly meshed) layer 2 network. The abstraction of the ETSI QKD key application interface (GS QKD 004) allows to get keys between any logical pair of devices which can and want to communicate without configuring the specific QKD node topology into the encryptors. Then the keys could be used in a hybrid way, combining the QKD key with a classical key exchange mechanism and using the already existing integrated classical key-management functions of the devices to keep all existing network functionality of the layer 2 encryptors. It was also important to keep the existing approved classical key management to minimize the cryptographic-relevant changes, ease a later approval of the QKD enhancement, and have a fallback to classical security in case of problems with the QKD modules. The payload data is then AES-256 encrypted on one or up to four 10 GbE interfaces.

## Optical Layout

The devices described in the last section were deployed in production facilities in Madrid (Fig. 2). Typically, several QKD systems were installed in each node and the low-level optical structure for the quantum and associated devices of the network is outlined in Fig. 3. Since it was a main target to explore how to include quantum communications in a standard production network, one of the guiding principles was to limit as much as possible the changes to the underlying standard optical network configuration. In this spirit, modifications were limited to add the mandatory multiplexers to add/drop the corresponding wavelengths and the already mentioned SDN-controllable optical switches to create additional quantum channels. The global control of the network was done through the SDN controller described above.

In any case, safety measures were taken to guarantee that the classical links were never disrupted. Backed by industrial grade solutions, these links could not be harmed by any of the tests in order to fulfil the strict service level agreements. In the end, after almost three years of testing many different use cases and configurations, it is interesting to highlight the performance, stability, and classical compatibility of these links, since it clearly shows how QKD technology has matured overtime and can be used jointly with classical communications.

Physically, the network has two different domains. The three rightmost nodes shown in Fig.3 are the Telefónica domain. This domain is connected with the RedIMadrid domain, with all the remaining nodes, through the border link connection (link 4: Quevedo-Norte). Since it is important for network operators to ensure that they can procure the equipment from different manufacturers and test their compatibility at the physical and logical level, the mixture of devices is larger in the Telefónica domain. In this domain all three manufacturers were present and the same physical media, a pair of optical fibers, was shared between the classical and QKD links.

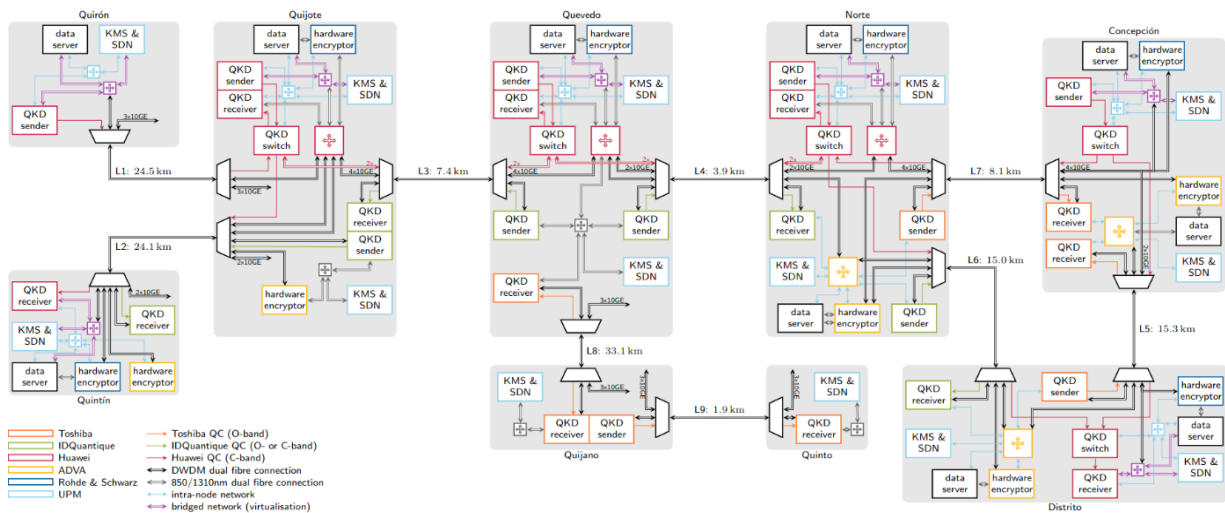

*Figure 3: Low-level optical structure of MadQCI testbed network. shows the optical set-up of the network including the quantum and directly associated equipment. The multiplexers and number of fibers are only shown indicative due to complexity reasons. The direct links follow the same numbering than in Fig. 2 and are numbered L1 through L9. (Li meaning link "i", with "i" the link number in the white labels in Fig. 2) The three rightmost nodes form a separate domain and are operated by Telefónica containing Norte, Concepción and Distrito. The Telefonica Norte site is connected to the Quevedo site in the RedIMadrid network over link 4 forming the border link. It is served by two QKD systems from two different manufacturers operating in different CV and DV modes. The longest link is link 8 between Quijano and Quevedo with a length of 33 km.*

The border link is served by two QKD pairs, one is DV (ID Quantique) and the other CV (HWDU). This critical link uses two technologically different QKD systems in the spirit of enhancing resilience and security by using a mixed configuration, where the final secret keys are obtained from two different sources.

The optical spectrum in one of the shared fibers in this link (direction Quevedo->Norte) is shown in Fig. 4 (left side). It presents the C-band spectrum (from 192 to 197 THz, 1521 to 1561 nm, approximately). The notch where the CV quantum channel is placed (within Ch37/Ch38) can be seen together with other three 10G telecom data channels (Ch21-Ch23) and the service channel for the DV-QKD system (Ch30). In this link, the DV quantum channel is located in the O-band. It is interesting to mention that moving to 100G data channels reduces the noise in the quantum channel, since the higher speed connections are more bandwidth efficient, and the optical power leaked out of its nominal wavelength is lower.

Figure 4 (right) shows the optical spectrum of link 8, the longest direct link in the network (33,1 km) that is served by a DV-QKD system provided by Toshiba. Similar to the QKD system of link 4, the quantum channel is in the O-band while all the data, including encrypted traffic, and service channels are in the C-Band. This reduces the noise in the quantum channel at the expense of higher losses (roughly 50% more in the O-band compared to the C-band). In the measured optical spectrum again only telecom channels are visible, although they include the service and key distillation channels serving the quantum link.

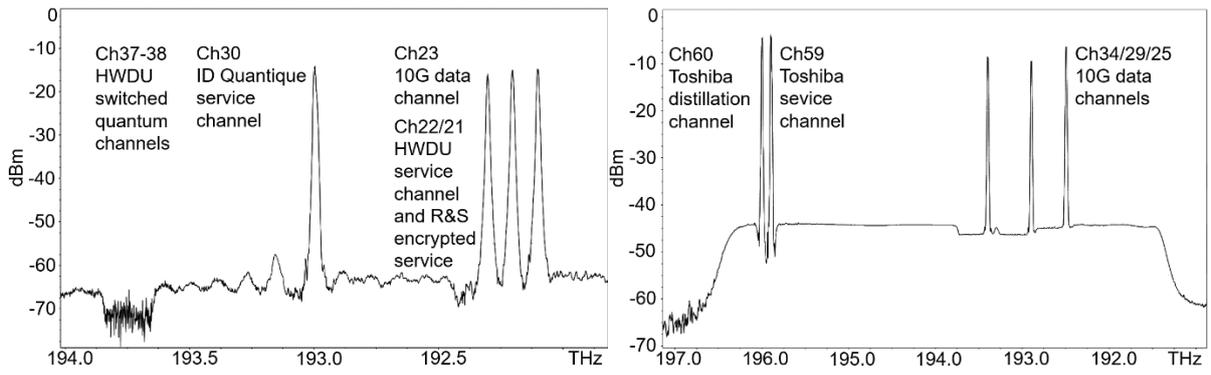

*Figure 4: The optical spectra of the border link number 4 (left) and the longest link 8 (right) are displayed. The OSA capture of link 4 in the C-band in the direction Quevedo->Norte shows three data channels (Ch21-Ch23) as well as the service channel of the ID Quantique system at Ch30 co-propagating with the quantum signals. Whereas the quantum channel of the ID Quantique system is in the O-band and therefore outside of the measurement range, the quantum channel of the HWDU device co-propagates in this fiber at the C-band (Ch 38 or 37, depending on the switched path) where a notch can be seen. On the right side, the C-band spectrum of link 8 (33.1km and 11.8 dB losses) is shown. In this link a DV quantum channel in the O-band co-propagates with the classical data channels (25-29-34), the service and distillation channels (59-60) in the C-band.*

To give a more detailed view on the performance in different links, Fig. 5 shows the QBER and secret key rate graphs that correspond to those discussed in Fig. 4 (Links 4 and 8 in Fig. 2 --or 3--). Link 8 (Fig. 5, right side) is the longest link in the network and is served by a DV (Toshiba) system. The quantum channel is in the O-band while all the data, including encrypted traffic, and service channels are in the C-Band. This reduces the noise in the quantum channel at the expense of about 50% higher losses. Note that the performance in both links varies significantly; although both are DV, the QKD protocol and basic parameters of the systems are quite different. One uses a COW protocol while the other is an optimized BB84 decoy states protocol. Note that the scale used for the key rate differs by an order of magnitude between both sides. Besides this, it was found that the Cerberis 3 generation of systems were very sensitive to variations in temperature, even those inside a datacenter, producing a secret key rate with more fluctuations than expected. QBER and secret key rate in these links are displayed for a period of about half a year, highlighting the long-term stability of the network. The performance, stability, and classical compatibility of these links clearly shows how QKD technology has matured over time.

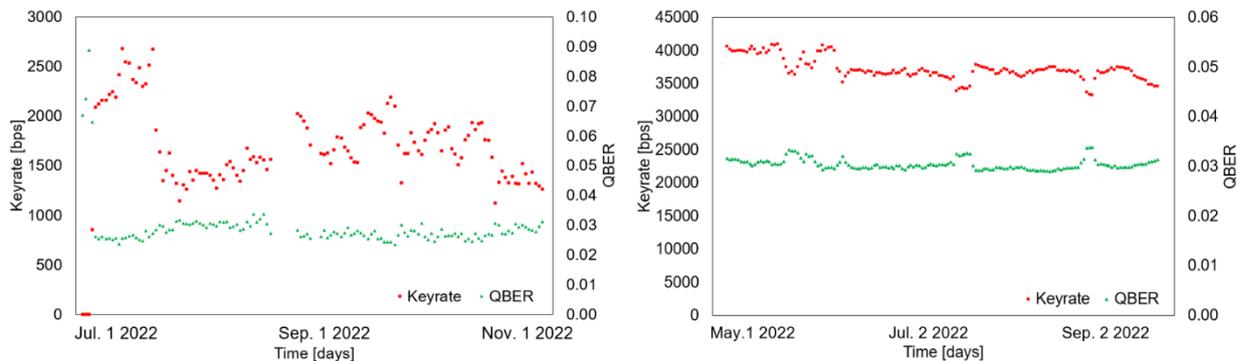

*Figure 5 Examples of measured QKD performances. (left) Link 4 (border link) The QBER and secret key rate over a period of about half a year is shown. Segments without data were due to maintenance labors. (right) Link 8 (longest link) QBER and secret key rates of the QKD system provided by Toshiba is shown for a similar period. The disruptions seen are due to maintenance work.*

In principle, quantum and classical signal coexistence, sharing the available optical fiber, was achieved in different ways in different links (see Table 1): CV-QKD systems are operated in a noise resilient regime, which allows to use the C-band for both, quantum and classical channels. DV-QKD systems have been usually operated with the quantum channel at the O-band, while keeping the classical channels in the C-band. The direction of the quantum channel coincided with the direction of the telecom operation. The integration of DV-QKD devices operating in the O-band was not evaluated systematically and independently by network operators so far. Within MadQCI, a comparison of different operation scenarios was possible.

Additionally, other configurations were also tested to address less standard situations. In Link 2 (Fig. 3), a DV-QKD link and CV-QKD link were operated together in the C-band on a single fiber for the quantum channels, while the second fiber in the pair was used for all data telecommunication channels. For the latter, bidirectional transceivers and multiplexers were used creating the upstream and downstream classical channels over the second fiber. This configuration is not very common, although it is supported by manufacturers. Interference among quantum channels was easily prevented by simple measures like avoiding the use of exactly the same wavelengths.

The logical compatibility and transparency towards the applications operated by users even from different domains was guaranteed by the software stack built for the network following the SDN mechanisms described above. At large, MadQCI has the focus of heterogeneous integration without restrictions to deploy hardware from different suppliers operated in different domains with the focus of smooth implementation of the SDN approach being transparent to the applications, that do not need to worry about the complexities of the underlying network. This was demonstrated through the implementation and testing of many different use cases.

## Use cases.

To underline the maturity of QKD technology and assess its feasibility, the network was running many qualitatively different use cases during the last three years. A short description of the most significant ones is provided in the supplementary material. A detailed description of the results is beyond the scope of this contribution and will be published elsewhere. This section is included here just to provide a general idea of the global performance from the network and applications point of view and the type of results obtained, given the broad range of applications tested.

Since it is the practical network usability what concerns us here, the traditional device-centric figures of merit presented in Fig. 1 (SKR, QBER, etc.) have a limited significance. With the architecture and redundancies available it is clear that there is enough key rate for various security applications that consume key, but that does not immediately translate in key indicators to judge the performance of an application, and other telecommunications metrics, closer to the use case are needed.

The use cases studied cover a wide range of services, from critical infrastructure protection, secure network management, cloud, 5G and final user services (e.g., e-health) as well as experimenting with new protocols, not related to direct encryption. The corresponding high-level application metrics can be very different, not only in performance numbers, but also in the significant magnitudes. In some use cases (5G or those related to real-time applications), latencies are the key figure, while for others, it could be the throughput, the maximum number of applications being served simultaneously, etc.

More than 85,000 use cases instances have been executed to gather these high-level, application-specific and network data to derive the metrics relevant to the particular cases. To illustrate the variety of this type of metrics, not commonly seen in research papers concentrating on the low-level performance of single links, but highly relevant for the network design and operation and also for the final user, we show several of them as examples in Table 2.

*Table 2: Metrics for different use-cases and sample values. 11 qualitatively different use-cases were tested in the network under many different situations. For each use-case, significant metrics were defined, and their value measured. Many of these metrics have additional dependencies on, for example, the specific node and how that node is configured, the number and type of QKD modules installed, losses in the links, etc. The table just intends to give a hint about how different these use-case metrics are compared with the basic per-link metrics usually published (e.g., QBER, key rate) in the literature. The values are just example values taken for some particular settings or node configurations.*

| Use-Case | Metric | Example Value | Use-Case | Metric | Example Value |
|---|---|---|---|---|---|
| OPoT | Increased Latency of processing a packet | +5.8 msec | e-health services (no real time/real time) | Encrypted data throughput / latencies | 500 Mbps / 1.3 msec. |
| Critical Infrastructure Protection | Throughput per user | 4Gbps (with 100 users) | B2B and 5G use-cases | Latencies in serving a request. | 0.15 msec |
| QKD as a cloud service | Number of requests served per second | 1230+-230 requests per second | Self-healed network management | Deployment time of software images | 24 sec. |

To be more precise in the illustration of the differences between the low-level QKD metrics and the application-level ones, we present a little more detail for a selected use case depicting the latencies in the Ordered Proof of Transit (OPoT) [23] use case. A detailed analysis of all the use cases is out of the scope of the present paper.

OPoT is a networking application that targets the problems of network security and attestation. The problem is to make sure that the data packets have passed through a defined set of nodes (e.g., a firewall) and in the correct order. The OPoT method used here to prove this requires QKD keys and, since the number of packets in the network can be huge, the processing time must be as low as possible. Hence, latency is a key figure of merit. The latencies with and without OPoT are presented in Fig. 4, showing that the approach is, indeed, feasible within a modern-day QKD network as the presented here.

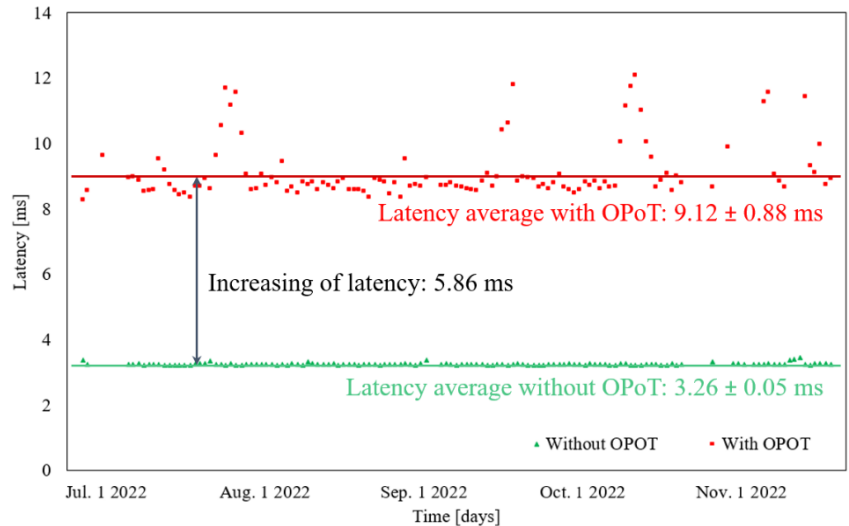

*Figure 6: Network metrics for the OPoT (Ordered Proof of Transit) use case (see text). In this case, the relevant metric is the latency. Each network packet is processed and tagged using a method based on QKD [17]. This serves the purpose to guarantee that the packets have passed through a specific set of nodes and in a specified order. Since the number of packets is very large, it is important to make sure that the increased latency penalty is limited. The plot shows that the average latency increase is about 5.86 ms for packets with OPoT (red), to be compared with the 3.26 ms without OPoT (green). The measures were taken during a period of a few days to demonstrate the stability of the metric during a long working period. The peaks typically correspond to times in which there was no key available due to the interaction with other use cases running simultaneously.*

# Conclusion

In the Madrid quantum network, in contrast to past efforts where the quantum part was a specially built, ad hoc network, we have developed a network to demonstrate the integration of QKD within production networking and security ecosystems. We believe that this integration, facilitating the access to quantum communications benefits as an easy to use, scalable service and where much infrastructure can be shared to avoid large up-front costs, is key to the development and broad adoption of QKD and, in general, quantum communications.

To this end we have used as a deployment base two already running, production-grade networks, with very little or no modification of their physical infrastructure to add quantum communications. The effort was put into the logic that glues the quantum and classical networks and enables them to share much of the infrastructure. To achieve this, we have used the software defined networking paradigm. This is a widespread paradigm in classical telecommunications networks that achieves its flexibility by decoupling data forwarding from control and management planes. QKD devices can become part of a specific forwarding plane, exporting its capabilities and requirements to a logically centralized SDN controller, where we have built the logic to control and enable quantum communications integrated within the classical network. In this way we can seamlessly add quantum capabilities to an existing network. This was demonstrated using approved standards and very heterogeneous QKD, optical transport and security hardware, together with a range of software tools and protocols well known by telecommunications companies. In this way, showing manufacturer independence, standardization, and common toolsets, we expect to create confidence and facilitate the widespread adoption of quantum communications technologies.

The capabilities of the network were demonstrated on a large set of nodes, many with several QKD modules installed, deployed using this new architecture. A typical security infrastructure based on standard hardware encryptors, and protocols was also implemented together with the network. QKD keys were also mixed with conventional keys from standard public-key cryptosystems, to achieve a better integration with the security ecosystem. A large set of use cases was run for testing and performance purposes, gathering low-level as well as application metrics. They were quite diverse, ranging from applications to secure critical infrastructures or network control to cloud and 5G applications. The testing took place over a period close to three years, where most of the nodes were continuously active. The results clearly show the feasibility of this approach to build large QKD networks.

MadQCI has brought together the highest number of different industrialized QKD links operated in a complex production network in Europe and during the largest period so far, showing that they can be integrated in the telecommunications and security infrastructures using modern network paradigms, tools, and standards. The network has demonstrated interoperability among manufacturers and network operators and a potential for scalability that can act as a blueprint with strong implications for a future European wide QKD network infrastructure.

## Acknowledgements

The authors would like to thank projects OpenQKD, EU H2020 grant 857156, Madrid Quantum – CM, funded by the European Union, NextGenerationEU (PRTR-C17.I1) and by Comunidad de Madrid, Programa de Acciones Complementarias, the EU Horizon Europe project "Quantum Security Networks Partnership" (QSNP), grant 101114043 and QuantERA II Programme, EU H2020 research and innovation program under Grant 101017733, with funding organizations: Foundation for Science and Technology – FCT, Agence Nationale de la Recherche - ANR, and Spanish Agencia Estatal de Investigación – AEI and EuroQCI-Spain, DEP grant 101091638

## AUTHOR CONTRIBUTIONS

V.M., D.L. and J.P.B conceived the project and designed the architecture. R.B-M, J. S-B. R.V developed the codes. D.R., F.P. and C.S. helped with the deployment in the RedIMadrid Network. M.P., H.H.B., F.F., A.P., F.F., A.J.S, R.I.W, H.G, S.R, F.I, C.A., M.H. provided the QKD systems, QRNGs, optical transport systems, encryptors and support. H.H.B., A.S-L, J.M. R-M., A. P-P. and J.F. were the main contributors to the optical design. V.M., M.P., A.P., L.O., H.H. and J.B.P. wrote the paper and analyzed the results with the help of the rest of the authors. All authors revised the manuscript. V.M. initiated and managed the collaboration.

## COMPETING INTERESTS

The authors declare no competing interests.

# Supplementary Material

Sept. 2023

### Abstract

This is a brief description of the use-cases run in the MadQCI network. It is intended to show the variety of applications. Note that the results presented are still tentative. A final analysis of the performance is still pending.

**Network security and attestation.** This use case sought to secure the operation, administration, and management (OAM) tasks of production networks. This was tested using an implementation of an advanced proof of transit (PoT) protocol, namely the ordered-PoT (OPoT), which ensured the transit of any packet of information through some functions in the network, so it can be tracked.

| Domains | TID, RM | Developer | UPM, TID |
|---|---|---|---|
| **Starting date** | 05.2020 | **End date** | 12.2022 |
| **Traffic delivered** | Attestated packets. | | |
| **Technology deployed** | SDN-like ordered proof of transit (OPoT). | | |
| **Sample results** | An increment of 5.86 ms in the packet processing latency when QKD is used. | | |

**Critical Infrastructure Protection.** The security of critical infrastructures, such as electrical and industrial installations, etc. was tested by providing them with a secure encryption method to protect typical industrial data traffic (SCADA).

| Domains | TID, RM | Developer | UPM |
|---|---|---|---|
| **Starting date** | 05.2021 | **End date** | 12.2022 |
| **Traffic delivered** | SCADA frames for industrial purposes. Level 1 and 2 encryption. | | |
| **Technology deployed** | SDN-aware IPSec software encryption. | | |
| **Sample results** | Up to 600 Mb/s encrypted traffic. | | |

**QKD as a Cloud Service**. This use case tested how the LKMS key manager would withstand multiple requests in a cloud environment where multiple client containers or virtual machines need to consume encryption key. Thus, both the number of co-existing connections and the amount of extracted key is measured.

| Domains | TID, RM | Developer | UPM |
|---|---|---|---|
| **Starting date** | 05.2021 | **End date** | 07.2022 |
| **Traffic delivered** | Simulated key requests from hosted VM. | | |
| **Technology Deployed** | Key management system of SD-QKD Stack. | | |
| **Sample results** | Up to 1000 requests supported. | | |

**e-Health services.** This use case used an QKD E2E encrypting technology for biomedical information to enhance the privacy and security of e-Health digital services.

| | | | |
|---|---|---|---|
| **Domains** | TID, RM | **Developer** | UPM |
| **Starting date** | 05.2021 **End date** | | 11.2022 |
| **Traffic delivered** | | Simulated medical information. | |
| **Technology deployed** | SDN-aware IPSec software encryption. Level 1 and 2 encryptors. | | |
| **Sample results** | Up to 600 Mb/s encrypted traffic. | | |

**Quantum Cryptography for B2B and 5G.** In this use case, the information encrypted by the implementation of the IPsec technology was carried over a 5G core infrastructure. This 5G link also hosted the e-Health use case.

| | | | |
|---|---|---|---|
| **Domains** | TID, RM | **Developer** | UPM, TID |
| **Starting date** | 05.2021 **End date** | | 11.2022 |
| **Traffic delivered** | | Simulated B2B transaction. | |
| **Technology deployed** | SDN-aware IPSec software encryption and simulated 5G core network. | | |
| **Sample results** | Up to 600 kb/s encrypted traffic. | | |

**Self-healed network management.** In this use-case, the aim was to enhance network resilience. The encrypted delivery of a virtual appliance to an OpenStack cloud deployment environment was tested. This environment can manage the lifespan of the virtualised resources with a security that nobody has modified during deployment.

| | | | |
|---|---|---|---|
| **Domains** | TID, RM | **Developer** | UPM |
| **Starting date** | 09.2021 **End date** | | 11.2022 |
| **Traffic delivered** | | NVF-ready virtual images. | |
| **Technology deployed** | Custom software encryption and Open Stack as the target software for virtual image deployment. | | |
| **Sample results** | 1 GB appliances delivered in 10 seconds. | | |

**Quantum Cryptography with minimal amount of QKD devices allowing independent protection of users in collocated computing centers**. The quantum channel switching capability and the ability to control it from the SDN controller with a global view, allows to allocate resources to protect the data traffic in a way that no other user can access the communications channel, not even in principle.

| | | | |
|---|---|---|---|
| **Domains** | TID, RM | **Developer** | HWDU, UPM |
| **Starting date** | 09.2020 | **End date** | 07.2022 |
| **Traffic delivered** | — | | |
| **Technology deployed** | SDN controlling by UPM for configuring optical capabilities of CV-QKD and optical network domain by HWDU. | | |
| **Sample results** | Up to 34 QKD links available with 10 QKD modules. | | |

**Security independence of a network provider from QKD device manufacturers**. Several quantum channels from different devices (and different manufacturers) were used to create a secret key independent among them for the same end-to-end link. The final key used is the combination of the several keys produced through the different QKD devices. A dishonest manufacturer can reveal the keys from their devices but has no knowledge of the keys of other manufacturers, thus the link will be still secure while at least one of the manufacturers is honest.

| | | | |
|---|---|---|---|
| **Domains** | TID, RM | **Developer** | UPM |
| **Starting date** | 09.2020 | **End date** | 07.2022 |
| **Traffic delivered** | Derived keys from two sources. | | |
| **Technology deployed** | SDN controlling and key managing by UPM for extracting key from multivendor sources (CV-QKD by HWDU and DV-QKD by IDQ). Encryptors at level 1 and 2 | | |
| **Sample results** | An increment of 1 ms in delivering latency when two QKD sources are combined. | | |

**Open Call QuGenome: Quantum Enabled Private Recognition of Composite Signals in Genome**. involved the collaboration with the partners Aveiro - Instituto de telecomunicacoes, from Portugal, for the provision of beyond-QKD services and Huawei. This Open Call addressed enabling private recognition of composite signals in genome and proteins. To do so, it needed to consume two different quantum resources, symmetric key and "raw" key. The former key material was the usual one distributed with QKD while the latest were the raw sequences resulting from the preparation and measurement of quantum states. In the Madrid test bed, the service access method based on the ETSI GS QKD 004 specification was modified to serve the two quantum resources interchangeably. Also, for serving the two resources simultaneously, the symmetric key was extracted from the IdQ partner's QKD systems and the raw key from the Huawei partner's one, using its optical switching mechanism. Both were deployed on a pair of links where the above-mentioned co-existence had been achieved.

| | | | |
|---|---|---|---|
| **Domains** | TID, RM | **Developer** | UPM, IT |
| **Starting date** | 12.2020 | **End date** | 02.2022 |
| **Traffic delivered** | Genomic data. | | |
| **Technology deployed** | Multi-party computation capabilities based on oblivious transfer by IT, SDN controlling and key managing by UPM for delivering symmetric and raw key from multi-vendor sources (CV-QKD by HWDU and DV-QKD by IDQ). | | |
| **Sample results** | Succesful multi-party computation of genomic data. | | |

**Open Call KaaS: Key as a Service.** It was developed in collaboration with the Up and Running Support Services partner, from Spain. It tested access to quantum resources by users who do not own QKD terminals but were within the practical security framework, such as trusted nodes. Thus, these users can obtain symmetric key from their respective node with an enhanced security scheme based on the pre-sharing of a master key that was used later for interacting with the ETSI GS QKD 004 access interface. This pre-sharing operation is based on using secure technologies, such as SSH, over a path diversity scheme that makes interception much more difficult. This use case was deployed and performed in the same scenario, namely Quintín, Quijote and Quevedo nodes.

| Domains | TID, RM | Developer | UPM, UAR |
|---|---|---|---|
| **Starting date** | 12.2020 | **End date** | 02.2022 |
| **Traffic delivered** | Quantum-distributed keys to final users. | | |
| **Technology deployed** | Multi-technology access network to QKD with path diversity by UAR and UPM. | | |
| **Sample results** | Succesful key distribution to final users. | | |

**OpenCall QGeKO:** Quantum Secure Distribution of Precise Global Navigation Satellite System Keys and Orders. To have a robust navigation capability available even in the case of deliberate interference (jamming) or signal spoofing is of paramount importance. The implementation of the capability of securing the navigation signals includes the ground distribution of keys for receiving the encrypted satellite signals, against any attack coming from malicious eavesdroppers that could use any kind of technology, including the quantum ones. And in some cases implies the use of the so called Secondary Channel, purposely created for this activity

| Domains | TID, RM | Developer | UPM, GMV |
|---|---|---|---|
| **Starting date** | 07.2022 | **End date** | 12.2022 |
| **Traffic delivered** | Global navigation satellite system (Galileo) Public Regulated Services secondary channel messages. | | |
| **Technology deployed** | PRS testbench system access software. | | |
| **Sample results** | Succesful GNSS data delivery through an encrypted channel. | | |